\documentclass[useAMS,usenatbib]{mn2e}
\usepackage{amsmath}
\usepackage{amssymb}
\usepackage{epsfig}
\title[Neutral hydrogen in galactic fountains]{Neutral hydrogen in galactic fountains}
\author[C. M. Booth and Tom Theuns]{C. M. Booth$^{1}$\thanks{E-mail: c.m.booth@durham.ac.uk (CMB)}, Tom Theuns$^{1,2}$\\
$^{1}$Institute for Computational Cosmology, University of Durham, South Road, Durham DH1 3LE\\
$^{2}$University of Antwerp, Campus Groenenborger, Groenenborgerlaan 171, B-2020 Antwerpen, Belgium\\}

\newcommand{\hi}{H$\,${\sc i}}

\newcommand{\apjl}{ApJL}

\begin{document}

\pagerange{\pageref{firstpage}--\pageref{lastpage}} \pubyear{2006}
\maketitle
\label{firstpage}

\begin{abstract}
  Simulations of an isolated Milky Way-like galaxy, in which
  supernovae power a galactic fountain, reproduce the observed
  velocity and 21~cm brightness statistics of galactic neutral
  hydrogen (\hi). The simulated galaxy consists of a thin \hi\ disk,
  similar in extent and brightness to that observed in the Milky Way,
  and extra-planar neutral gas at a range of velocities due to the
  galactic fountain. Mock observations of the neutral gas resemble the
  \hi\ flux measurements from the Leiden-Argentine-Bonn (LAB) \hi\,
  survey, including a high-velocity tail which matches well with
  observations of high-velocity clouds. The simulated high-velocity
  clouds are typically found close to the galactic disk, with a
  typical line-of-sight distance of 13~kpc from observers on the solar
  circle. The fountain efficiently cycles matter from the centre of
  the galaxy to its outskirts at a rate of around $0.5 M_{\odot} {\rm
  yr}^{-1}$.
\end{abstract}


\section{Introduction}
\label{sec:intro}

Radio observations of the 21~cm hydrogen emission line reveal that the
Milky Way (MW) contains a thin \hi\ disk surrounded by a population of
\lq clouds\rq\, with velocities incompatible with models of galactic
rotation yet apparently not part of the Hubble flow either
(\cite{mull63}). The nature of these \lq high-velocity clouds\rq\,
(HVCs) remains somewhat unclear mostly because it is difficult to
determine their distances and hence infer physical properties.

Observations of stars with known distances along the line of sight to
a HVC can be used to constrain the distance to the cloud, by testing
whether or not it is detected in absorption in the stellar spectrum
(see e.g. \cite{schw95}). Unfortunately such constraints are available
for only a relatively small number of HVCs (\cite{wakk01}).
\cite{putm03} use constraints from H$\alpha$ emission, and find that
most HVCs are within $\sim 40$~kpc, except for those associated with
the Magellanic stream. \cite{brun01} claim that some fraction of HVCs
display a head-tail morphology, may be a result of interaction with a
diffuse ambient galactic wind (\cite{quil01}). Searches for stars
associated with HVCs have so far resulted in non-detections
(e.g. \cite{hopp07}). Metallicities have been measured for a small
number of HVCs and vary over a wide range (e.g. \cite{gibs01}),
suggesting that HVCs are not a homogeneous set.  Finally one can
observe other galaxies to infer the nature of HVCs from their
projected distances. The Andromeda galaxy (M31) has a population of
HVCs close ($\le 50$~kpc) to its disk (\cite{west07}). Other nearby
spiral galaxies also contain extra-planar neutral gas (see
e.g. \cite{barb05}). \cite{pisa07} have searched for HVC analogues in
six loose groups of galaxies, similar to the Local Group. Their
failure to detect compact HVCs implies that any \hi\ clouds are near
to the galaxies in this group: they are not roaming freely throughout
the group itself.  These observations do not unambiguously determine
the nature of HVCs, and several theoretical models for them have been
proposed. Oort (1970) discusses a model where the HVCs are close to
the MW disk, and result from of a \lq galactic fountain\rq\
(\cite{shap76}). Gas from the galactic disk rises buoyantly after
being heated by supernovae (SNe), becomes thermally unstable and cools
radiatively into neutral clouds. Once the clouds are dense they fall
back ballistically onto the disk, and are seen as the HVCs.  Whether
SNe explosions can start a galactic fountain has been investigated
both theoretically (\cite{kahn98}) and through numerical simulations
(\cite{deav98,deav01}).

Blitz et al. (1999; see also \cite{brau99}) suggest (some) HVCs are
neutral gas associated with the numerous dark matter substructures
seen in simulations of haloes of galaxies and groups of galaxies. Such
HVCs are at large distances ($\gtrsim$ 40~kpc) from the galactic
centre, and distributed throughout the local group. \cite{conn06}
report the detection of \hi\, clouds in fully
cosmological $\Lambda$CDM simulations of galaxy formation, and their
HVCs are reminiscent of Blitz's clouds. \cite{mall04} discuss how a
population of clouds at distances of $\sim 150$~kpc may result from
the halo gas becoming thermally unstable.  Given the rich variety HVC
properties it seems likely that more than one of the above models,
with an additional component resulting from tidal interactions
(e.g. \cite{ferg06}) is required to explain the full cloud
distribution.  In this letter we perform numerical simulations of star
formation in an isolated Milky Way-like galaxy, in which SNe power a
galactic fountain. We analyse the simulation in terms of mock \hi\
observations and demonstrate they reproduce many of the observed
features of the \hi\ disk and its HVCs.

This letter is structured as follows.  In section \ref{sec:numerical}
we provide details of the numerical scheme used in our investigation
and in section \ref{sec:results} we present a comparison between our
simulations and the distribution of galactic \hi.

\section{Numerical Method}
\label{sec:numerical}

\subsection{Star Formation and Feedback model}
\label{sec:sf}
The star formation and feedback prescriptions used in the simulations
is described in detail in \cite{boot07} (hereafter BTO07), here we
briefly review its main features. The scheme treats the interstellar
medium (ISM) in terms of three distinct but interacting components:
cold ($T\lesssim 10^2$~K) molecular clouds surrounded by a warm
($T\sim 10^4$~K) ambient phase in the disk, interspersed by a hot
($T\sim 10^6$K) tenuous phase that extends into the halo, and is
powered by SNe. Gas is cycled through these phases by a number of
processes.

Conversion of gas to stars is regulated by the rate at which molecular
clouds form and get destroyed. Clouds form from thermally unstable
ambient gas and get destroyed directly by feedback from massive stars,
and indirectly through thermal conduction. Gas cooling is due to
Compton and line cooling, using interpolation tables generated using
{\sc Cloudy}\footnote{We thank our colleagues J Schaye, C Dalla
Vecchia and R Wiersma for allowing us to use these routines.}
(\cite{ferl98}). Feedback in the ambient gas phase cycles gas into a
hot galactic fountain or wind. We model the ambient and hot gas phases
hydrodynamically using smoothed particle hydrodynamics (SPH,
\cite{ging77,lucy77}) as implemented in the {\sc Gadget2} simulation
code (\cite{spri05}). Motivated by the fact that we cannot resolve the
Jeans mass of the molecular gas, the clouds are modelled using \lq
sticky particles\rq\ that move ballistically through the ambient gas,
but may coagulate when colliding. When such a cloud has built-up
enough mass through accretion and coagulation to form a Giant
Molecular Cloud (GMC), it collapses into stars, which then destroy the
GMC through stellar winds and SNe explosions.

BTO07 show that this model produces a multiphase medium with cold
clouds, a warm disk, hot SN bubbles and a hot, tenuous halo, similar to
that observed in spiral galaxies. The star formation rate, surface \hi\
density, molecular fraction, and molecular cloud mass-spectrum of
simulated galaxies match closely those observed in the MW.  

\subsection{Model Galaxy}
\label{sec:galaxyics}
The initial conditions for our MW-like galaxy are generated using a
publicly available programme, {\sc GalactICs} (\cite{kuij95}), which
generates a near equilibrium galaxy consisting of an approximately
exponential disk, a bulge and a (dark matter) halo, using
collisionless particles. The bulge, disk and halo have masses in the
ratio 0.31:1.00:28.27 (giving a baryon fraction of $\sim$0.3 times the
universal value). The total mass of the system is $2.21\times 10^{12}
M_{\odot}$ and the circular velocity at the solar radius, $r=8.5$~kpc,
is $\sim 220 {\rm km\,s}^{-1}$.

We convert 10\% of the disk particles into SPH particles with
temperature $10^4K$, the remaining 90\% of the disk represents a
stellar disk with population ages set by assuming that the galaxy had
a constant star formation rate of $1M_{\odot}{\rm yr}^{-1}$.  The
bulge is changed in its entirety into an old stellar population.
Finally 1\% of the material in the halo is changed into SPH particles
with a temperature of $10^6K$. The galaxy is now no longer in
equilibrium and starts converting atomic gas into molecular clouds,
which in turn form stars, yet soon settles into a near-equilibrium
state where feedback regulates the star formation rate at
approximately 1$M_{\odot}/{\rm yr}$. It was shown in BTO07 that the
resulting galaxy resembles the MW in terms of its gas fraction,
molecular fraction, gas distribution and Schmidt-Kennicutt star
formation law. Our highest resolution simulation initially contains
$\sim 3\times 10^6$ particles in the disk (particle mass $\approx
1.5\times 10^5M_\odot$), but it has been demonstrated in BTO07 that
that the properties of this galaxy are not strongly dependent on
resolution.

\subsection{Mock \hi\ observations}
We assume that the \hi\ fraction of the ambient gas is set by the
balance between photo-ionisation, collisional ionisation, and
recombinations, which we compute using {\sc Cloudy} (\cite{ferl98}),
imposing the UV background at redshift $z=0$ given by \cite{haar01}.
This \hi\ gas can be detected by its hyperfine emission line
(\cite{vand45}). A hydrogen atom in its ground state has higher energy
when the spins of proton and electron, $J_{\rm p}$ and $J_{\rm e}$,
are parallel than when they are anti-parallel. Spontaneous transitions
to the lower energy state occur at a rate governed by the Einstein
coefficient, $A_{10}=2.85\times 10^{-15}$~s$^{-1}$, and result in the
emission of a \lq 21~cm photon\rq\ of wavelength $\lambda=21.11{\rm
cm}=c/\nu$. For the densities, $n_{\rm H}$, and temperatures, $T$,
relevant to the ISM, the typical time $\tau$ between collisions, $\tau
\sim 100 \times ({n_{\rm H}/ {\rm cm}^{-3}})^{-1}\,({T/10^4{\rm
K}})^{-1/2}{\rm yr}$, is much shorter than the spontaneous 21cm
transition rate, $\tau_{1/2}=A_{10}^{-1}\,{\rm s}\sim 10^7\,{\rm yr}$,
and so collisions keep the levels of the low and high states, $n_0$
and $n_1$, in the ratio $n_1=3n_0$. The mean emission rate of 21~cm
photons per neutral hydrogen is then given by $A_{10}n_1=(3/4)A_{10}$.
The 21~cm flux of a cloud with neutral hydrogen mass $M_{\rm HI}$ at
distance $D$ is
\begin{eqnarray}
\label{eq:flux}
F &=& {\frac{(3/4)\,A_{10}\,(M_{\rm HI}/m_{\rm H})\,h\nu}{\,4\pi\,D^2}}\nonumber\\
  &\approx & 2\times 10^{-19}\, \frac{M_{\rm HI}}{M_\odot}\,\Big(\frac{1\,{\rm kpc}}{D}\Big)^2\,\hbox{\rm erg cm$^{-2}$ s$^{-1}$\,,}
\end{eqnarray}
(see e.g. \cite{spit78}).  In radio astronomy, this is quoted in terms
of the line flux, $\int S(\nu)\,dv$, with $S(\nu)$ expressed in Jy and
the line-width in km s$^{-1}$, $F=\int S(\nu)\,d\nu=\int
S(\nu)\,(\nu/c)\,dv$, hence (\cite{wakk91b})
\begin{eqnarray}
\label{eq:snudnu}
\int \Big[\frac{S(\nu)}{{\rm Jy}}\Big]\,\,\Big[\frac{dv}{{\rm km s}^{-1}}\Big] &=& \frac{1}
 {{\rm Jy}}\, \frac{1}{{\rm km s}^{-1}}\frac{c}{\nu}\,F\\
 &=&\frac{1}{0.235}\frac{M_{\rm HI}}{M_\odot}\,\Big(\frac{{\rm kpc}}{D}\Big)^2\,.
\end{eqnarray}
The 21cm \lq brightness temperature\rq\,,$T_B$, of an object is defined
as the temperature at which a black-body emits the same flux. The
conversion from flux to brightness temperature is then given by
$T_B/S=R$, where the telescope-dependent conversion factor we use is
$R=0.158 {\rm K\,Jy}^{-1}$ in order to match the observational survery
we are comparing our results against (\cite{huls88}).  To compute the
simulated \hi\ flux we place an observer in the simulated galaxy and
evaluate the net flux received by an ideal radio telescope with a beam
size $\theta$. For a single SPH particle at position ${\bf r}_i$, the
fraction $dm/m$ of mass that falls within the beam at distance between
$r$ and $r+dr$ is
\begin{equation}
dm/m= \bigg(2\pi\,r^2 (\cos(\theta)-1)\, dr\bigg)W(|{\bf r}_i-{\bf r}|,h)\,,
\end{equation}
where $h$ is the smoothing length of the particle, and $W$ the SPH
kernel.  The total flux received from this particle is computed from
Eq.~(\ref{eq:flux}), and is represented by a Gaussian emission line
centred at velocity ${\bf v}\cdot{\bf r}/r$ with width
$\sigma_v=(k_{\rm B}\,T/m_h)^{1/2}$. The total spectrum is obtained
integrating over $dr$ and summing over all particles.

Simulations have been performed at two different mass resolutions to
ensure that resolution effects do not affect our results. Observations
have been repeated for a number of observers along the solar circle,
and at times spanning a period of 1Gyr. We use these observers to
compute error bars on the mock observations.  The distribution of
\hi\, at galactic latitudes, $b$, greater than 20$^\circ$ does not
depend strongly on the time at which observations are made, showing
that the galaxy our results do not represent a transient feature.
Within the galactic disk ($b<20^\circ$) the mean brightness
temperature of the gas decreases slowly over time as the gas disk is
converted into stars.

\section{Results and Discussion}
\label{sec:results}
\vspace{-2mm} 
\begin{figure*}
\begin{center}
\centerline{\scalebox{0.9}{\includegraphics[width=\textwidth,clip,viewport=-40 8 750 340]{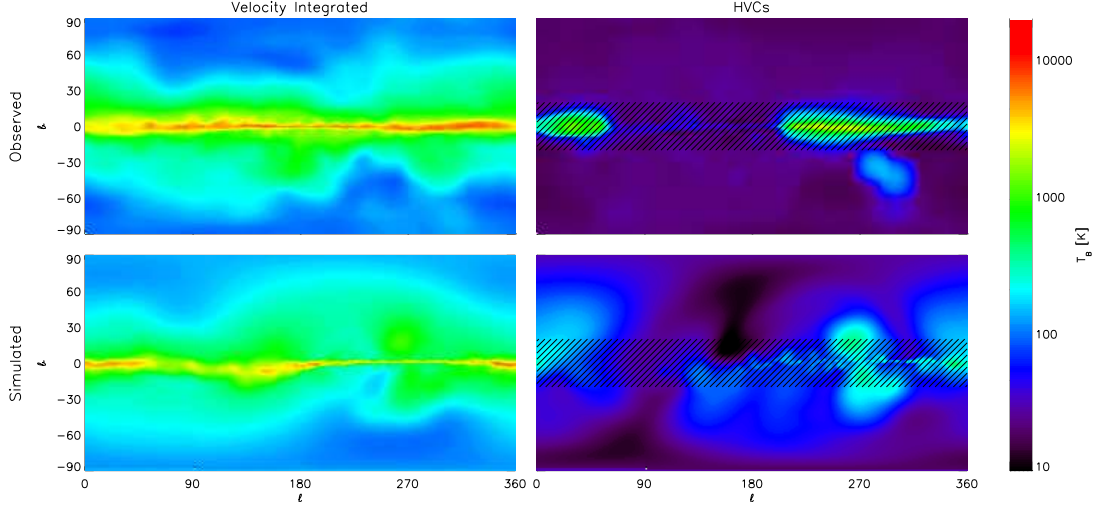}}}
\end{center}
\caption{All-sky maps of 21~cm brightness temperatures in the LAB
  survey (top panels) and in simulations (lower panels), integrated
  over all velocities (left panels) or integrated over high velocities
  ($|v_{\rm lsr}| > 100$~km s$^{-1}$) only (right panels).  The
  observational data has been convolved with a spatially adaptive
  Gaussian, to make the angular resolution of both the simulated and
  observed data the same.  Both observed and simulated galaxy consist
  of a thin ($b< 5^\circ$) \hi\ disk with brightness temperature $T_B>
  5000$~K, embedded in a thicker ($b\sim 45^\circ$) disk with $T_B\sim
  1000$~K, with $T_B\sim 100$~K at higher latitudes. The high velocity
  gas has $T_B\lesssim 100$~K in both observed and simulated
  galaxy. The simulation does not reproduce the smaller-scale
  structure seen in the LAB due to lack of numerical resolution.\vspace{-5mm}}

\label{fig:allsky}
\end{figure*}

The all-sky 21~cm brightness distribution of the simulated galaxy
looks remarkably similar to the observed \hi\ in the Milky Way, as
measured by the Leiden-Argentine-Bonn (LAB, \cite{kalb05}) \hi\,
survey (Fig.~\ref{fig:allsky}). The LAB survey has angular resolution
of $0.5^\circ \times 0.5^\circ$ but unfortunately our numerical
simulation does not have enough particles to resolve structures on
such small scales. We calculate the mean angular extent of the
particles that contribute to the flux in a given direction on the sky
in the simulated galaxy, and then smooth the LAB survey with a
Gaussian kernel of the same width. At low galactic latitudes this
smoothing angle is typically less than $1^\circ$.  However, at high
latitudes the mean smoothing length is much larger, and at
$|b|>60^\circ$ can reach up to $20^\circ$ due to the relatively small
number of SPH particles at these latitudes.

Both observed and simulated brightness maps display a bright and thin
\hi\ disk in the plane of the MW, embedded in a thicker cooler
envelope ($T_B\sim 10^3$~K), with an even cooler component ($T_B\sim
100$~K) at high velocities, $|v_{\rm lsr}| > 100$~km s$^{-1}$, with
respect to the local standard of rest (LSR). The brightness
temperature $T_B$, and its fall-off with latitude, is very similar in
the observed and simulated maps. The minimum \hi\, brightness
temperature in the simulated map is 95K, in good agreement with \hi\,
observations of the MW, where it is found that every line-of-sight
contains easily observable \hi. The simulated high velocity gas
($|v_{{\rm lsr}}|>100$~km s$^{-1}$) forms a nearly uniform background.
Due to the relatively small number of particles that are flagged as
high velocity at any one time the spatial resolution is poor,
especially at high latitudes, and these simulations do not resolve the
fine structure seen in the LAB.  When the LAB data are smoothed to the
same resolution as the simulation the resulting distribution matches
closely, with a mean brightness temperature of $T_B\sim 23.22$~K, as
compared to 20.33~K in the simulations.

\begin{figure}
\begin{center}
\includegraphics[width=8.3cm]{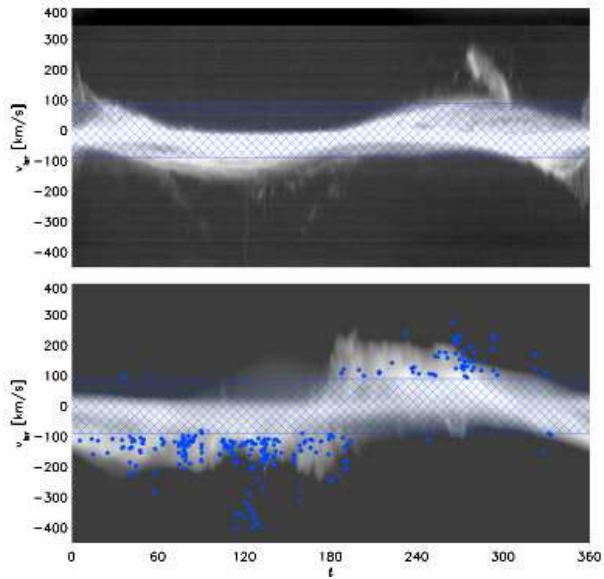}
\end{center}
\caption{Velocities with respect to the local standard of rest of \hi\
  gas in the LAB survey (top panel) and the simulated galaxy (bottom
  panel) as function of galactic longitude, $l$.  Intensity on this
  plot is the 21cm brightness of the \hi\,, integrated over galactic
  latitude. The simulated galaxy has \hi\ gas out to velocities of
  $|v_{\rm lsr}| \sim 200$~km s$^{-1}$ in a characteristic pattern
  also seen in the LAB. The blue points (bottom panel) are HVCs from
  the catalogue of Lockman et al (2002). The simulation does not
  resolve small clouds, but the velocity distribution of the simulated
  \hi\ traces the same regions in $(l,v_{\rm lsr})$ space as the
  observed clouds.\vspace{-5mm}}
\label{fig:veldist}
\end{figure}

The velocity distribution of the simulated \hi\, also matches well
with the LAB data (Fig \ref{fig:veldist}). Although we cannot resolve
individual HVCs, the simulated \hi\, velocities are in good agreement
with the properties of the HVC catalogue of
\cite{lock02}. \cite{lock02} identified some of the detections in this
survey with external galaxies, we have removed these from the
plot. Additionally, clouds that were identified as being part of the
Magellanic stream, which dominate the extreme negative velocity flow
(\cite{math74}), were removed.

\begin{figure*}
\centerline{\scalebox{0.9}{\includegraphics[width=\textwidth,clip]{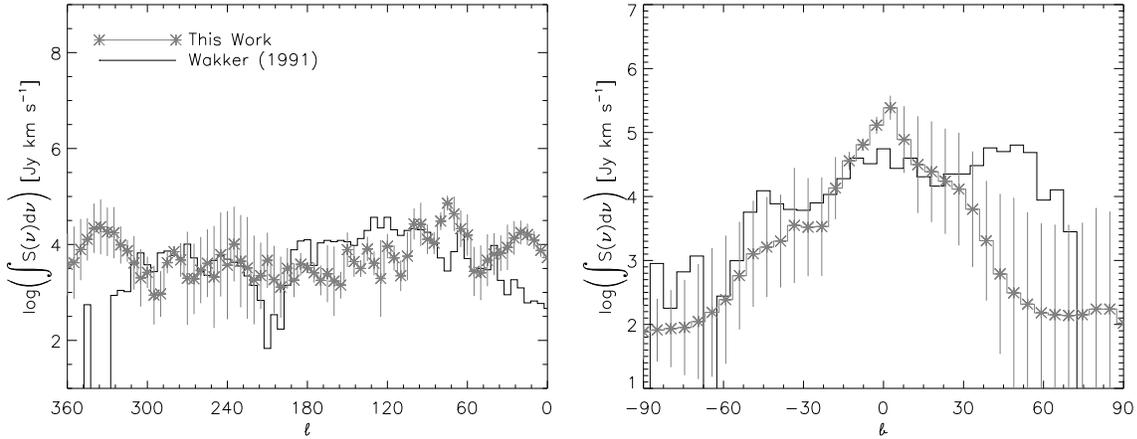}}}
\caption{Line flux of 21~cm emission in galactic longitude $l$ (left
  panel) and latitude $b$ (right panel) for high velocity gas. The
  solid black lines are from the survey for HVCs by Wakker (1991).
  The observational data has the populations of clouds belonging to
  the Magellanic stream and outer arm removed, as an isolated MW
  galaxy can not reproduce these. The grey histogram is obtained from
  the simulations, with error bars a measure of the scatter between
  simulated observers at the solar circle. The simulated galaxy looks
  very similar to the observed one in $l$, but the distribution in $b$
  is slightly more extended in the real MW, although the simulated
  scatter is large.\vspace{-5mm}}
\label{fig:hist}
\end{figure*}

The line flux, $\int S(\nu)\,d\nu$, for high velocity gas, $|v_{{\rm
lsr}}|>100 {\rm \,km\,s}^{-1}$, is shown in Fig.~
\ref{fig:hist}. Solid black lines represent the results of
\cite{wakk91} with the emission due to the Magellanic stream and outer
arm of the MW removed, as we do not expect an isolated galaxy to match
these features. Errors on the mock data are calculated by repeating
the measurements for observers at different points along the solar
circle. The mock and real data look remarkably similar. The
distribution of neutral gas perpendicular to the galactic plane is
approximately exponential with a scale height of 5~kpc, in agreement
with the predictions of \cite{breg80} for a galactic fountain.
Although as noted in Fig \ref{fig:hist} the distribution of 21cm
emission in galactic latitude is slightly more concentrated in the
simulated data than the observed data, when averaged over the whole
sky the mean values of the smoothed maps agree to within 10\% and are
485K for the observed data, and 446K for the simulated data.

\begin{figure}
\begin{center}
\includegraphics[width=7.6cm,clip,viewport=0 0 400 290]{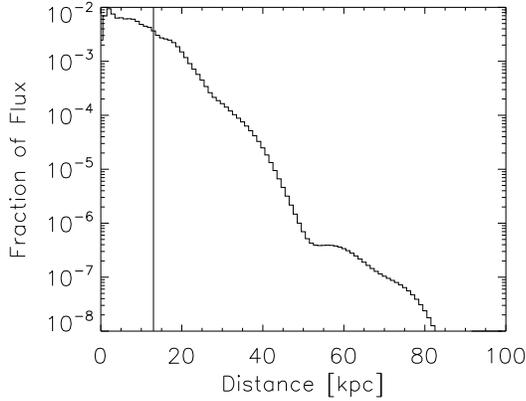}
\end{center}
\caption{Flux weighted mean distance to \hi\, emitting high velocity
  ($|v_{{\rm lsr}}|>100 {\rm \,km\,s}^{-1}$) gas in the simulated
  galaxy. Fifty per cent of the emission occurs within $\sim 13$~kpc
  (vertical line).\vspace{-5mm}}
\label{fig:dist}
\end{figure}

It is difficult to measure the distances to HVCs in the real
universe. However, in our simulations this information is preserved
and can be measured easily (Fig \ref{fig:dist}). 50\% of the flux is
emitted within a distance of 13~kpc from our observer. This figure is
in line with that observed other galaxies (\cite{barb05,pisa07}),
notably in M~31 by \cite{west05}, who found that HVCs were generally
at a projected distance of less than 15~kpc.

\begin{figure}
\begin{center}
\includegraphics[width=7.6cm,clip,viewport=30 5 480 330]{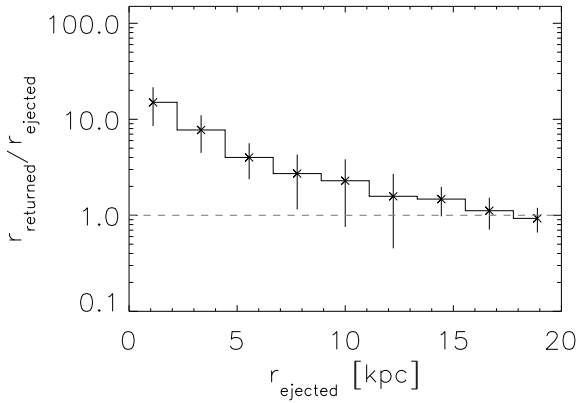}
\end{center}
\caption{Histogram showing the relation between the galactic radius at which gas is ejected from the galaxy ($r_{{\rm ejected}}$), and the radius at which it next passes through the galactic disk ($r_{{\rm returned}}$).  Overall the galactic fountain acts to move gas from the centre of the galactic disk to its outskirts.\vspace{-5mm} }
\label{fig:rad}
\end{figure}

Gas in the fountain rains back on the disk at a radius ($r_{\rm
returned}$) which is generally larger than the radius ($r_{\rm
ejected}$) from where it was launched (Fig \ref{fig:rad}).  An SPH
particle is classified as being part of the fountain if it is ejected
to a vertical distance of more than 2~kpc from the galactic plane and
at a later time falls back within 1~kpc of the galactic disk.  The
galactic fountain has an overall effect of moving gas from the inner
part of the galactic disk to its outskirts: most fountain particles
with $r_{\rm ejected}<8$~kpc have $r_{\rm returned}/r_{\rm
ejected}\gtrsim 4$ hence rain back outside the solar
circle. \cite{corb88} note that the effect of this \lq outer galactic
fountain\rq\, depends upon the physical properties of the gas that
returns to the disk at large radii and could either evaporate the
\hi\, disk, or cause it to grow. The rate at which gas returns to the
outer disk in the simulations, $\dot M\sim 0.5M_{\odot}$ yr$^{-1}$, is
in agreement with observational estimates and could be enough to drive
the observed turbulence (\cite{sant07}). We find that a peak mass flux
of at least $0.5M_{\odot}$ yr$^{-1}$ through the disk out to a radius
of 15kpc, suggesting that a galactic fountain of this form is capable
of driving ISM turbulence all the way to the outskirts of the galactic
disk.  In our simulations, $\dot M$ decreases gradually as the
quiescent star formation rate gradually decreases due to the depletion
of the local gas supply.  However, we note that in a fully
cosmological setting bursts of star formation due to galaxy mergers
coupled with infall of gas from the surrounding intergalactic medium
could continue to drive this process over the lifetime of the galaxy.

We conclude that our multi-phase star formation implementation
naturally produces a galactic fountain in a Milky Way-like model
galaxy. The neutral hydrogen in the fountain has a spatial and velocity
distribution in good agreement with a variety of observations for the
MW and other spiral galaxies.
\section*{Acknowledgements}
\vspace{-2mm} 
CB thanks PPARC for the award of a research studentship.

\label{lastpage}

\end{document}